\documentclass[12pt]{article}

\usepackage{multirow}
\usepackage{comment}
\usepackage{epsfig}
\usepackage{epstopdf}
\usepackage{amssymb,amsmath}
\usepackage{comment}

\usepackage{setspace}

\newcommand{\bea}{\begin{eqnarray}}
\newcommand{\eea}{\end{eqnarray}}

\begin{document}
\begin{titlepage}
%
%
\vspace*{10mm}
\begin{center}
\baselineskip 25pt 
{\Large\bf
$SU(5)  \times U(1)_X$ Axion Model with \\Observable Proton Decay  
}
\end{center}
\vspace{5mm}
\begin{center}
{\large
Nobuchika Okada\footnote{okadan@ua.edu}, 
Digesh Raut\footnote{draut@udel.edu}, 
and Qaisar Shafi\footnote{qshafi@udel.edu}
}
\end{center}
\vspace{2mm}

\begin{center}
{\it
$^{1}$ Department of Physics and Astronomy, \\ 
University of Alabama, Tuscaloosa, Alabama 35487, USA \\
$^{2,3}$ Bartol Research Institute, Department of Physics and Astronomy, \\
 University of Delaware, Newark DE 19716, USA
}
\end{center}
\vspace{0.5cm}
\begin{abstract}
We propose a $SU(5) \times U(1)_X \times U(1)_{PQ}$ model, where $U(1)_X$ is the generalization of the $B-L$ (baryon minus lepton number) gauge symmetry and $U(1)_{PQ}$ is the global Peccei-Quinn (PQ) symmetry. 
There are four fermions families in $\bf{{\overline 5}} + \bf{10}$ representations of $SU(5)$, a mirror family in $\bf{5}+\bf{{\overline {10}}}$ representations, and three $SU(5)$ singlet Majorana fermions. 
The $U(1)_X$ related anomalies all cancel in the presence of the Majorana neutrinos. 
The $SU(5)$ symmetry is broken at $M_{GUT} \simeq (6-9)\times 10^{15}$ GeV and the proton lifetime $\tau_p$ is estimated to be well within the expected sensitivity of the future Hyper-Kamiokande experiment, $\tau_p \lesssim 1.3 \times 10^{35}$ years. 
The $SU(5)$ breaking also triggers the breaking of the PQ symmetry, resulting in axion dark matter (DM), with the axion decay constant $f_a$ of order $M_{GUT}$ or somewhat larger. 
The CASPEr experiment can search for such an axion DM candidate.   
The Hubble parameter during inflation must be low, $H_{inf} \lesssim 10^9 $ GeV, in order to successfully resolve the axion domain wall, axion DM isocurvature and $SU(5)$ monopole problems.  
With the identification of the $U(1)_X$ breaking Higgs field with the inflaton field, we implement inflection-point inflation, which is capable of realizing the desired value for $H_{inf}$. 
The vectorlike fermions in the model are essential for achieving  successful unification of the SM gauge couplings as well as the phenomenological viability of both axion DM and inflation scenario.
\end{abstract}
\end{titlepage}

\newpage

\section{Introduction}
\label{sec:Intro}
A variety of well-established experimental results in particle physics and cosmology have exposed some of the inadequacies of the Standard Model (SM) of particle physics \cite {Zyla:2020zbs}. 
These include the confirmation of the existence of non-baryonic dark matter (DM), observation of tiny but non-zero masses for SM neutrinos, the observed asymmetry between the matter and antimatter abundance in the universe, the necessity of cosmic inflation in the very early stages of the universe's evolution, and the strong CP puzzle. 
The SM must be supplemented with new physics to account for these observations.

Among the various proposed extensions of the SM, the models based on grand unified theory (GUT) are attractive because they predict unification of the SM gauge interactions and also explain the quantization of the electric charges of the SM fermions \cite{GUT}.
An interesting grand unification scenario utilizes the anomaly free gauged $U(1)_X$ extension of the SM, where the $U(1)_X$ symmetry \cite{Appelquist:2002mw}  is the generalization of the $B-L$ (baryon minus lepton number) symmetry \cite{mBL}.  
The generalized $U(1)_X$ charge of each particle is defined as a linear combination of its hypercharge ($Q_Y$) and $B-L$ charge ($Q_{B-L}$), $Q_X= x_H \, Q_Y+ Q_{B-L}$, where $x_H$ is a free parameter \cite{Oda:2015gna}. 
For $x_H = -4/5$ \cite{Okada:2017dqs}, the SM quarks and leptons are unified in the ${\bf {\overline 5}}$ and ${\bf 10}$ representations of $SU(5)$. 
The three SM singlet Majorana neutrinos needed to cancel all the $U(1)_X$ related anomalies can explain the origin of observed neutrino masses and flavor mixings via the type-I seesaw mechanism \cite{Seesaw}. 
The unification of the three SM gauge couplings can be achieved by adding components of vector-like quark pairs from the ${\bf {5}} \oplus {\bf {\bar 5}}$ and ${\bf 10} \oplus {\bf {\bar 10}}$ representations of $SU(5)$ \cite{Okada:2017dqs}.

In this article, we propose a model based on  the symmetry $SU(5) \times U(1)_X \times U(1)_{PQ}$, where $U(1)_{PQ}$ is the global Peccei-Quinn (PQ) symmetry \cite{Peccei:1977hh}, which addresses all the inadequacies of the SM discussed above. 
The PQ symmetry solves the strong CP problem \cite{Peccei:2006as} and the associated axion from the PQ symmetry breaking is the DM candidate \cite{Weinberg:1977ma}. 
The $SU(5)$ symmetry breaking also triggers the breaking of $U(1)_{PQ}$, and so the DM physics is intimately connected to the physics of grand unification. 
In particular, the axion decay constant $f_a$ is comparable to the $SU(5)$ GUT symmetry breaking scale, $M_{GUT} \sim 10^{15}- 10^{16}$ GeV. 
To resolve the $SU(5)$ GUT monopole problem \cite{monopole1} one may consider the low scale inflation scenario with $H_{inf} \ll M_{GUT}$, where $H_{inf}$ is the value of the Hubble parameter during the inflation. See also Ref.~\cite{Kibble:1982ae}.  
However, in this case, the axion DM scenario suffers from the cosmological fatal axion domain wall problem and axion DM isocurvature problem  (for a review see, for example, Ref.~\cite{Kawasaki:2013ae}). 
With the axion decay constant $f_a \simeq M_{GUT}$, the resolution of the axion domain wall and axion DM isocurvature problems require a low value for $H_{inf} \lesssim 10^9 $ GeV \cite{Okada:2020cvq}.  
Well-known inflationary scenarios with the Coleman-Weinberg or Higgs potential with minimal coupling to gravity \cite{Shafi:2006cs},  and a quartic potential with non-minimal coupling to gravity \cite{Okada:2010jf} predict a relatively large $H_{inf} \simeq 10^{13-14}$ GeV \cite{Okada:2014lxa}.
With the identification of the $U(1)_X$ Higgs field with the inflaton field, we implement the so-called inflection-point inflation (IPI) scenario \cite{Okada:2016ssd}, which can realize $H_{inf} < 10^9 $ GeV. 
The new fermions in the model are key to achieving successful unification of the SM gauge couplings as well as the phenomenological viability of both the axion DM and the IPI inflation scenario.  
The Majorana fermions generate the observed baryon asymmetry via leptogenesis \cite{Fukugita:1986hr}. 
We identify sets of model parameters such that the new physics scenarios discussed above including proton decay are phenomenologically viable.

\begin{table}[t]
\begin{center}
\begin{tabular}{|c|ccc|c|c|c|}
\hline
                                 		 &SU(5)             &U(1)$_X$      &U(1)$_{PQ}$		         
\\ 
\hline
$\psi_{\overline {5}}^{i}$     	  &${\overline {\bf 5}}$       &$-$3/5      &$0$            
\\
$\psi_{10}^{i}$      			  &${\ {\bf 10}}$               &$+1/5$      &$0$                    
\\
\hline
${\widetilde \psi}_{5}$        				&${ {\bf 5}}$    			&$+$3/5             &$1$    
\\
${\widetilde \psi}_{\overline {10}}$               &${\overline {\bf 10}}$     &$-1/5$     		    &$1$		          
\\
\hline
$(N^{c})^{j}$        &{\bf 1 }    &$+1$  	  &$0$       
\\
\hline
$\Sigma$         & {\bf 24 }    &$0$          	  &$-1$
\\
$S$                & {\bf 1 }      &$0$          	  &$-1$         
\\  
$\Phi$             &{\bf 1 }       &$-2$           &$0$
\\
$H$                & {\bf 5 }      &$-2/5$         &$0$         
\\
\hline
\end{tabular}
\end{center}
\caption{
Particle content of $SU(5) \times U(1)_X \times U(1)_{PQ}$ model. 
It includes four fermion families, $\psi_{{\overline 5} (10)}^i$ ($i=1,2,3,4$), 
one mirror family, $\widetilde \psi_{{5} ({\overline 10})}$, 
three Majorana fermions, $(N^c)^{j}$  ($j=1,2,3$), and four complex scalars ($\Sigma$, $S$, $H$ and $\Phi$). All the fermions are in their left-handed spinor representation and $``c"$ denotes charge conjugation. 
}
\label{tab:1}
\end{table}

\section{Model}
\label{sec:model}
The particle content is listed in Table~\ref{tab:1}. 
The model includes four fermion families, $\psi_{{\overline 5} (10)}^i$ ($i=1,2,3,4$) in the ${\overline {\bf 5}} ({\bf 10})$ representations of $SU(5)$, 
one mirror family, $\widetilde \psi_{{5} ({\overline {10}})}$ in the ${\bf 5} ({\overline {\bf 10}})$ representation, 
three $SU(5)$ singlet Majorana fermions, $(N^c)^{j}$  ($j=1,2,3$), and four complex scalars ($\Sigma$, $S$, $H$ and $\Phi$).  
All fermions are in their left-handed spinor representations and $``c"$ denotes charge conjugation. 
The $U(1)_X$ related anomalies cancel in the presence of $N^c$'s.  
Since only the new fermions are charged under the PQ symmetry, this model may be regarded as the $SU(5) \times U(1)_X$ realization of the Kim–Shifman–Vainshtein–Zakharov (KSVZ) axion model\footnote{The symmetry group of the GUT model discussed in Ref.~\cite{Okada:2020cvq} is also $SU(5) \times U(1)_{PQ} \times U(1)_X$. However, there are two key differences between the two models: (i) The Higgs sector in Ref.~\cite{Okada:2020cvq} includes two Higgs doublets and is effectively a Dine-Fischler-Srednicki-Zhitnitsky (DFSZ) type axion model \cite{Dine:1981rt}. See also Ref.~\cite{FileviezPerez:2019ssf}. (ii) The $SU(5)$ adjoint field in Ref.~\cite{Okada:2020cvq} is a singlet under PQ symmetry, and so there is no direct connection between the $SU(5)$ and PQ symmetry breaking scales.}. See also Ref.~\cite{FileviezPerez:2019fku}.

In the following, we consider the spontaneous breaking of  $SU(5)$ and $U(1)_{PQ}$ symmetries, followed by the breaking of $U(1)_X$ and the electroweak symmetry. 
The Higgs potential for $\Sigma$ and $S$ fields is given by 
\bea  
V (\Sigma, S) &=& 
 - \mu_\Sigma^2 {\rm Tr}[ \Sigma^{\dagger}\Sigma]
+ \lambda_{1} \left( {\rm Tr}[ \Sigma^{\dagger}\Sigma]\right)^2
+ \lambda_2 {\rm Tr}[ \Sigma \Sigma ] {\rm Tr}[ \Sigma^{\dagger}\Sigma^{\dagger} ]
+ \lambda_3 {\rm Tr}[ \Sigma^{\dagger} \Sigma \Sigma^{\dagger} \Sigma ]
+ \lambda_4 {\rm Tr}[ \Sigma^{\dagger}\Sigma^{\dagger} \Sigma  \Sigma ]
\nonumber \\
&&
-\kappa_{1} \left({\rm Tr}[ \Sigma^{\dagger}\Sigma^{\dagger} \Sigma ] S  +{\rm h.c}\right)
+\kappa_{2}  {\rm Tr}[ \Sigma^{\dagger} \Sigma ] \left(S^{\dagger} S \right)
-\kappa_{3} \left( {\rm Tr}[ \Sigma^{\dagger} \Sigma^{\dagger} ] S^2+{\rm h.c.}\right)
\nonumber \\
&&
-\mu_S^2 \left(S^\dagger S\right)  +\lambda_S \left(S^\dagger S\right)^2.  
\label{eq:SSPot}
\eea
Here, the couplings parameters are all real and positive and we have neglected mixed terms between $\Sigma$ and $S$ fields with $H/\Phi$, assuming that the associated couplings to be adequately small because $H/\Phi$ fields not essential for the breaking of $SU(5)$ and $PQ$ symmetries. 
The $SU(5)$ and $U(1)_{PQ}$ symmetry breaking is accomplished by the $\Sigma$ and $S$ fields vacuum expectation values (VEVs), namely,  
$\langle\Sigma\rangle = v_\Sigma/(2\sqrt{15}) \;{\rm diag} (-2,-2,-2,3,3)$ and $\langle S\rangle  = v_S/\sqrt{2}$. 
Solving the stationary conditions for the potential in Eq.~(\ref{eq:SSPot}) we obtain    
\bea
\mu_\Sigma^2 &=& \frac{1}{60} \left(-3 \sqrt{30} \kappa_1 v_S v_{\Sigma }+30 \left(\kappa_2-2 \kappa_3\right) v_S^2+2 \left(30 \lambda _1+30 \lambda _2+7 \left(\lambda _3+\lambda _4\right)\right) v_{\Sigma }^2\right),
\nonumber \\
\mu_S^2 &=& \frac{1}{60v_S} \left(30 \kappa_2 v_S v_{\Sigma }^2-60 \kappa_3 v_S v_{\Sigma }^2+60 \lambda _S v_S^3-\sqrt{30} \kappa_1 v_{\Sigma }^3\right). 
\label{eq:mincond}
\eea 
Applying these results to evaluate the mass spectrum for the scalar and gauge fields we obtain $12$ superheavy massive gauge bosons, 
37 massive scalars and one massless scalar. 
The details about the scalar mass spectrum is presented in Appendix 1. 
For concreteness, let $v_{\Sigma }=  \sqrt{30} \kappa_{1} v_S$, with the coupling parameters $\kappa_{1,2} = \lambda_1= \lambda_S = 0.3$, $\kappa_3 = -0.011$, $\lambda_2 = -0.049$, and $\lambda_3 = \lambda_4  = 0.375$ such the 35 scalar masses are given by
\bea
m_1^{(1)} &=& 0.20 v_S, 
\qquad 
m_2^{(3)} = 0.35 v_S
\qquad
m_3^{(8)} =  0.42 v_S
\qquad
m_4^{(8)} = 0.71 v_S
\nonumber \\
m_5^{(3)} &=& 0.58 v_S
\qquad
m_6^{(12)} = 0.39 v_S
\qquad
m_7^{(1)} = 0.57 v_S
\qquad
m_8^{(1)} = 1.1 v_S, 
\label{eq:NSmass}
\eea
where the numbers in the exponents are the degeneracy of each mass eigenvalue. 
The massless scalar field, which we identify to be the axion, is given by 
\bea
a (x) = \frac{1}{\sqrt{v_S^2 + v_\Sigma^2}} \left(v_S \; \chi_S (x) + v_\Sigma \; \chi_\Sigma (x) \right), 
\eea
where $\chi_S$ ($\chi_\Sigma$) is the imaginary component of $S$ ($\Sigma$) that acquires the VEV.

Following $SU(5) \times U(1)_{PQ}$ symmetry breaking, the residual symmetry is $SU(3)_c \times SU(2)_L \times U(1)_Y \times U(1)_X$.  
In the following, we neglect the mixing between $\Phi$ and $H$ Higgs fields which will be justified later. 
It allows us to independently examine the $\Phi$ and $H$ sector Higgs potential.  
The VEV of $\Phi$ far exceeds the electroweak VEV of $H$, and so the $U(1)_X$ symmetry is primarily broken by $\Phi$. 
Setting 
\bea
\Phi(x) &=& \frac{1}{\sqrt{2}}\left(\phi(x) + v_{X}\right) e^{i \chi(x)/ v_{X}}, 
\label{eq:phi} 
\eea 
where $v_X$ denotes its VEV, $\Phi$ potential is given by 
\bea
V(\Phi) =  \lambda_\phi \left(\Phi^\dagger \Phi  - \frac{v_{X}^2}{2}\right)^2. 
\label{eq:PotP}
\eea
The breaking of the $U(1)_X$ symmetry by the VEV of $\Phi$ also generates masses for the $U(1)_X$ gauge boson $Z^\prime$ and the real component $\phi$, 
\bea 
m_{Z^\prime} = 2 g v_{X}, \; \; 
m_\phi = \sqrt{2 \lambda_\phi} v_{X}, 
\label{eq:mass}
\eea
respectively, where $g$ is the $U(1)_X$ gauge coupling. Finally, the electroweak symmetry gets broken after the charge neutral component of the $SU(2)_L$ doublet Higgs field in $H$ field acquires its VEV, $v_{H} = 246 $ GeV.

Let us now consider fermion masses. 
We introduce the Yukawa interactions only for ${\psi}_{{\overline 5}(10)}^4$ and ${\widetilde \psi}_{{5} ({\overline {10}})}$, 
\bea
{\cal L}  &\supset& 
-  {\psi}_{\overline {5}}^4 \left( {\tilde y}_5^4 S - {\widetilde Y}_5^4 \Sigma \right) {\widetilde \psi}_{5}
- {\psi}_{10}^4 \left( {\tilde y}_{10}^4 S - {\widetilde Y}_{10}^4 \Sigma \right) {\widetilde \psi}_{\overline {10}}. 
\label{eq:LY1}
\eea  
Because there is one copy of ${\widetilde \psi}_{5 ({\overline {10}})}$, only one linear combination of the four ${\psi}_{{\overline 5}(10)}^i$ obtain a non-zero mass from the $S$ and $\Sigma$ VEVs. 
Here, without loss of generality we work in a basis where ${\psi}_{{\overline 5}(10)}^4$ and ${\widetilde \psi}_{{5} ({\overline {10}})}$ are the massive states. 
The decomposition of the pairs under the SM gauge group and their masses will be discussed in Sec.~\ref{sec:GCU}.

The Yukawa interactions of the fermions with $H$ are given by
\bea
{\cal L}  &\supset& 
-\sum_{i,j=1}^{4} Y_{H1}^{ij}H^\dagger \psi_{\overline 5}^i \psi_{10}^j 
- \sum_{i,j=1}^{4} Y_{H2}^{ij}H\psi_{10}^i \psi_{10}^j.  
\label{eq:LY2}
\eea 
In the following analysis we assume $Y_{H1, H2}^{i 4}\ll 1$ ($i = 1,2,3$), so that $\psi_{{\overline 5}, 10}^i$ ($i= 1,2,3$) are identified with the SM fermions and the mixing between the SM fermions and ${\psi}_{{\overline 5}(10)}^4$ is non-zero but negligibly small. 
This is crucial to ensure the decay of the exotic heavy fermions. 
The mass spectrum of ${\psi}_{\overline {5},10}^4$ will be discussed in Sec.~\ref{sec:GCU}.

The Yukawa interactions involving the Majorana neutrinos are expressed as  
\bea  
   {\cal L} \supset 
- \sum_{i=1}^{4} \sum_{\beta=1}^{3} Y_D^{i \beta} H \psi_{\overline 5}^i \left(N^c\right)^\beta
- \left(\frac{1}{2} \sum_{\beta=1}^{3} Y_M^\beta  \Phi {\left(N^c\right)}^{\beta} \left(N^c\right)^{\beta}  +{\rm h.c.}\right),
\label{eq:LY2} 
\eea 
where we have used the mass basis for ${\widetilde \psi}_{\overline {5}}^i$ and a flavor-diagonal basis for the Majorana neutrinos. 
After breaking of the $U(1)_X$ and the electroweak symmetry, the first and second terms in Eq.~(\ref{eq:LY2}) generate the Dirac and Majorana type masses for the neutrinos 
\bea 
m_D^{i \beta}= \frac{Y_D^{i \beta}}{\sqrt{2}} v_{H}, \; \; 
m_{N^\beta}= \frac{1}{\sqrt{2}} Y_M^\beta  v_{X}. 
\label{eq:masses}
\eea

\section{Gauge Coupling Unification}
\label{sec:GCU}
\begin{figure}[t!]
\begin{center}
\includegraphics[scale=0.9]{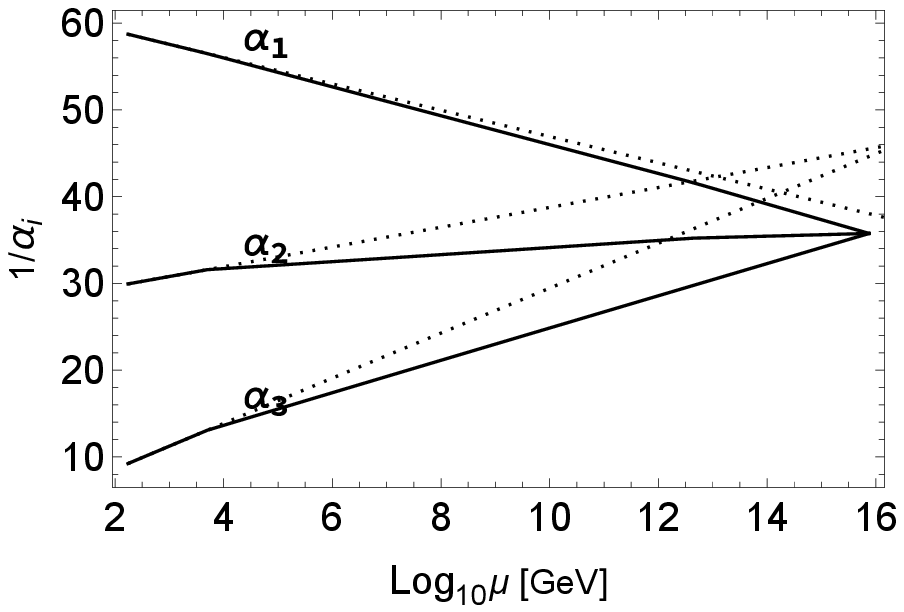}\;
\includegraphics[scale=0.9]{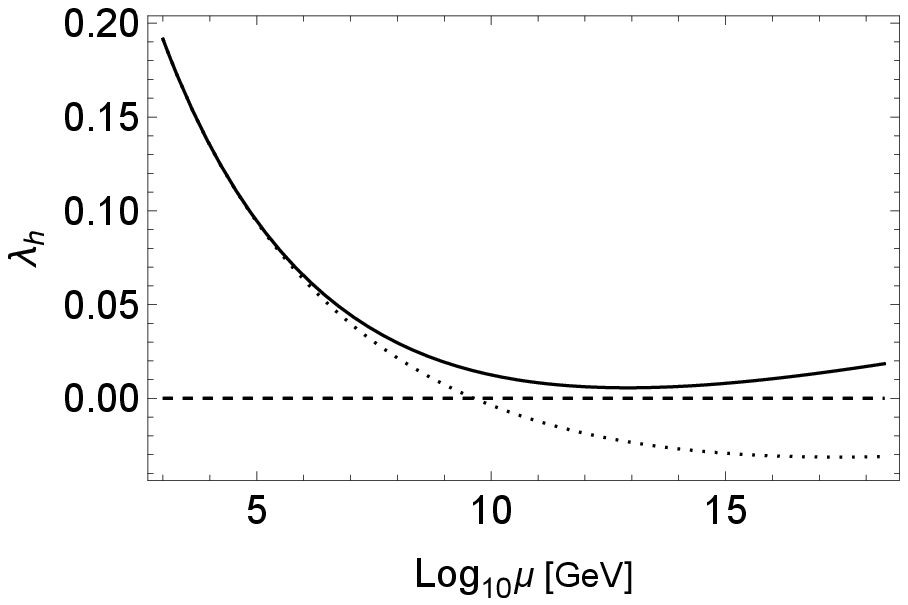}
\end{center}
\caption{
For fixed masses of the new lepton doublet $L$ and quarks ($D^c$ and $Q$) with masses $4.5\times10^{12}$ GeV and $5000$ GeV, respectively, the plot shows the RG running of SM couplings. In the left panel the diagonal solid (dotted) lines labeled $\alpha_i = g_i^2/4\pi$ ($i=1,2,3$) depict the SM $U(1)_Y$, $SU(2)_L$ and $SU(3)_c$ gauge couplings with (without) new fermions, respectively, which are unified at $M_{\rm GUT} \simeq 7.54 \times 10^{15}$ GeV. 
In the right panel, the solid (dotted) curve depicts the RG running of the SM Higgs quartic coupling with (without) vector-like fermions along with the horizontal dashed line depicting $\lambda_h = 0$.  
}
\label{fig:GCU}
\end{figure}
In this section we evaluate the renormalization group (RG) running of the SM gauge couplings including the contribution from the new fermions and scalars which have masses much smaller than the $SU(5)$ symmetry breaking scale $v_\Sigma$. 
For the benchmark values used in the previous section, all the new scalars in Eq.~(\ref{eq:NSmass}) have masses close to $v_\Sigma$. 
The new fermions in ${\psi}_{{\overline 5}(10)}^4$ and ${\widetilde \psi}_{{5} ({\overline {10}})}$ obtain their masses from the $S$ and $\Sigma$ VEVs in Eq.~(\ref{eq:LY1}):  
\bea
{\cal L}_{\rm mass}  &\supset& -{\psi}_{\overline {5}}^4 \left( {\tilde y}_5^4 \langle S\rangle - {\widetilde Y}_5^4 \langle \Sigma \rangle \right) {\widetilde \psi}_{5}
-  {\psi}_{10}^4 \left( {\tilde y}_{10}^4 \langle S \rangle - {\widetilde Y}_{10}^4 \langle \Sigma \rangle \right) {\widetilde \psi}_{\overline {10}}, 
\label{eq:LYnew}
\eea
The decompositions of ${\widetilde \psi}_{\overline {5},10}^4$ under the SM gauge group are  given by 
\bea
{ \psi}_{\overline {5}}^4 &=& D^c ({\bf 3^*, 1, -2/3}) \oplus L ({\bf 1, 2, -1/2}), 
\nonumber \\
{ \psi}_{10}^4 &=& U^c ({\bf 3^*, 1, -2/3}) \oplus Q ({\bf 3, 2, 1/6}) \oplus  E^c ({\bf 1, 1, 1}). 
\label{eq:decomp}
\eea
The SM decomposition of their partners, ${\widetilde \psi}_{5, \overline {10}}$, are the conjugate of the representations shown in Eq.~(\ref{eq:decomp}).  
Using the benchmark $v_{\Sigma }=  \sqrt{30} \kappa_{1} v_S$, we evaluate masses of the vector-like pairs within the multiplets.  
For ${\cal O} (1)$ Yukawa coupling values, we find that the pairs in ${\psi}_{{\overline 5}(10)}^4-{\widetilde \psi}_{{5} ({\overline {10}})}$ may have a large mass splitting between them. 
For example, if we fix the mass of $D^c$ ($Q$) to be ${\cal O} (1)$ TeV, the masses of the remaining components in the multiplet, without loss of generality, is approximately given by $ {\widetilde Y}_{10 (5)}^4 \times v_\Sigma$ . 
The CMS collaboration for the LHC has set the lower limit of around $1500$ GeV \cite{Sirunyan:2019ofn} at $95\%$ confidence level for vector-like quarks with hypercharge ($-2/3$) and the vector-like leptons doublets with hypercharge ($-1/2$) in the mass range of $120 - 790$ GeV \cite{Sirunyan:2019ofn} are excluded at $95\%$ confidence level.

In the following analysis of the RG running of the SM gauge couplings, let us fix the Yukawa couplings such that the lepton doublet $L$ has mass $M_L =4.5\times 10^{12}$ GeV, the quarks $D^c$ and $Q$ have degenerate mass $M_Q = 5000$ GeV, and $U^c$ and $E^c$ have GUT scale masses.   
We numerically solve the RG equations for SM gauge couplings listed in Appendix.~2. 
The left panel of Fig.~\ref{fig:GCU} shows our results for the RG running of the SM gauge couplings as a function of the energy scale $\mu$. 
The solid lines labeled by $\alpha_i = g_i^2/4\pi$ ($i=1,2,3$) denote the SM gauge couplings for $U(1)_Y$, $SU(2)_L$ and $SU(3)_c$, respectively. 
For comparison, in Fig.~\ref{fig:GCU} we also show the RG running of the SM gauge couplings in the absence of the new fermions which are depicted by the dotted lines. 
In the former case, the SM gauge couplings successfully unify at around $M_{\rm GUT} \simeq 7.53 \times 10^{15}$ GeV with the unified coupling value $\alpha_{GUT} = \alpha_1 = \alpha_1 = \alpha_3 \simeq 1/35.8$. 
Using these values, the proton lifetime from its decay mediated by the $SU(5)$ GUT gauge bosons can be approximated as \cite{Nath:2006ut}\footnote{For the discussion on the effects of threshold corrections on gauge coupling unification and proton decay estimate, see, for example, Ref.~\cite{Chakrabortty:2019fov},}
\bea
\tau_p \approx \frac{1}{\alpha_{GUT}^2} \frac{M_{GUT}^4}{m_p^5} \approx 9.39 \times 10^{34} \; {\rm years}, 
\label{eq:PD}
\eea  
where $m_p = 0.983$ GeV is the proton mass. 
This is consistent with the current experimental lower bound on proton lifetime given by the Super-Kamiokande with $\tau_p(p \to \pi^0 e^+) \gtrsim 4.0\times 10^{34}$ yr \cite{Miura:2016krn}. 
Importantly, the predicted lifetime is within the expected sensitivity reach of future Hyper-Kamiokande, $\tau_p \lesssim 1.3 \times 10^{35}$ yr \cite{Abe:2011ts}. 
The color triplet scalar field contained in $H$ can also mediate proton decay; the Super-Kamiokande  experiments excludes the colored scalar mass lighter than ${\cal O} (10^{11})$ GeV \cite{Nath:2006ut}. 
The validity of the proton lifetime estimate in Eq.~(\ref{eq:PD}) requires the colored Higgs mediated proton decay to be suppressed, particularly, the colored Higgs mass must to be greater than $10^{11}$ GeV. 
Consider the quartic interactions of $H$ with $S/\Sigma$ fields, for instance, $H^\dagger \Sigma^\dagger \Sigma H$. 
Since both $S$ and $\Sigma$ have VEVs close to $M_{GUT}$, consistency of proton lifetime estimate require these (positive) quartic couplings to be greater than ${\cal O}(10^{-11}$).

In the right panel of Fig.~\ref{fig:GCU} we show the RG running of the SM Higgs quartic coupling $\lambda_h$ as a function of the energy scale $\mu$. 
The solid (dotted) curve depicts the RG running with (without) the new fermions and the horizontal dashed line denotes $\lambda_h = 0$. 
With $\kappa _h(\mu) > 0$ for all values of $\mu$,  the SM Higgs potential is stabilized in the presence of the new fermions.

\begin{figure}[th!]
\begin{center}
\includegraphics[scale=0.9]{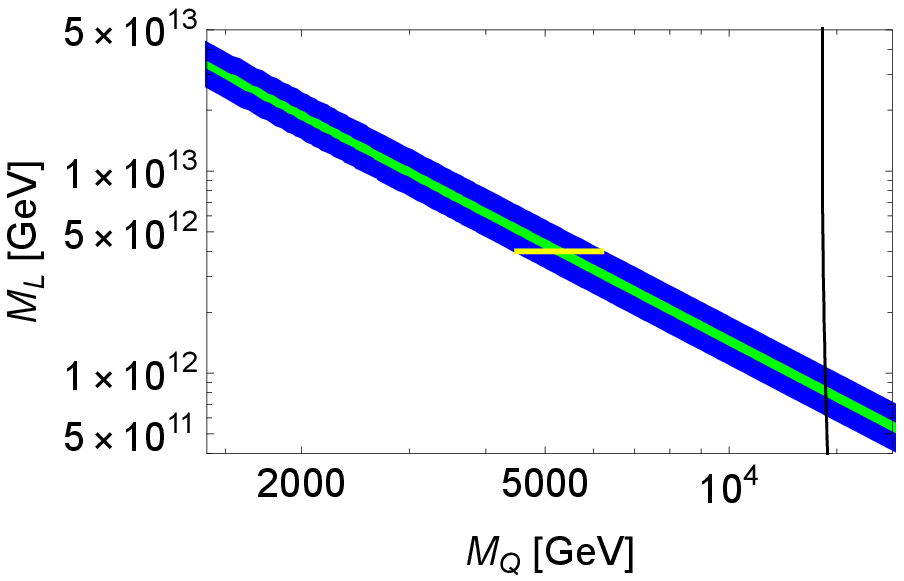}\;
\includegraphics[scale=0.9]{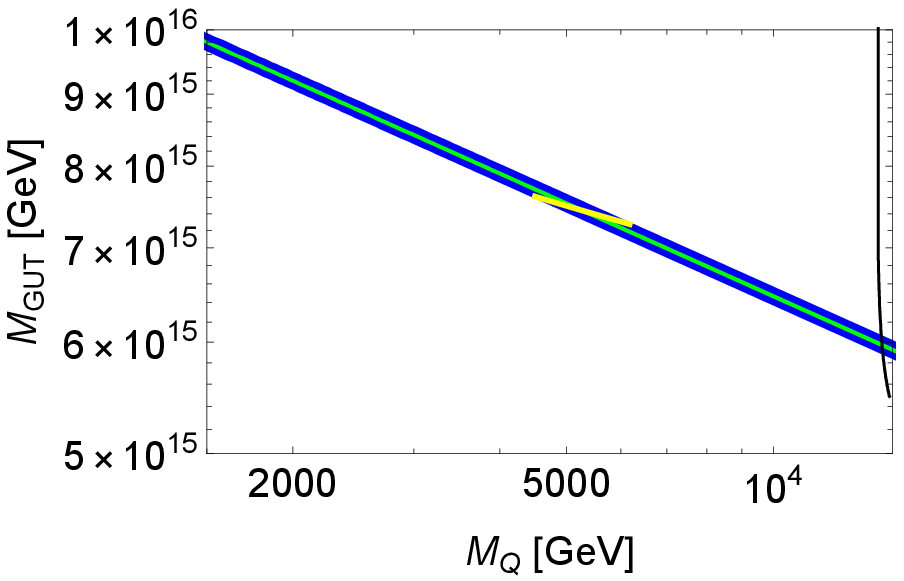}\\
\includegraphics[scale=0.9]{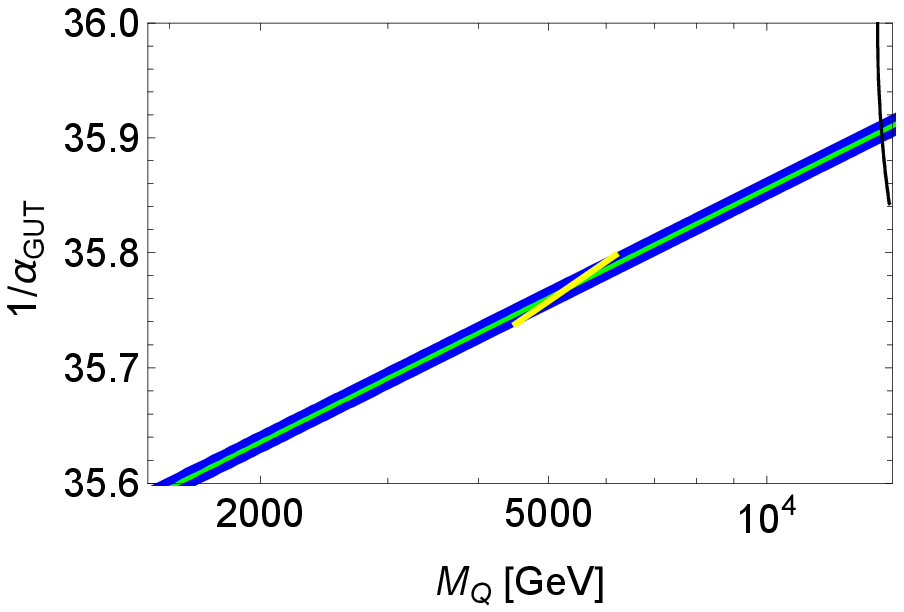}\;
\includegraphics[scale=0.9]{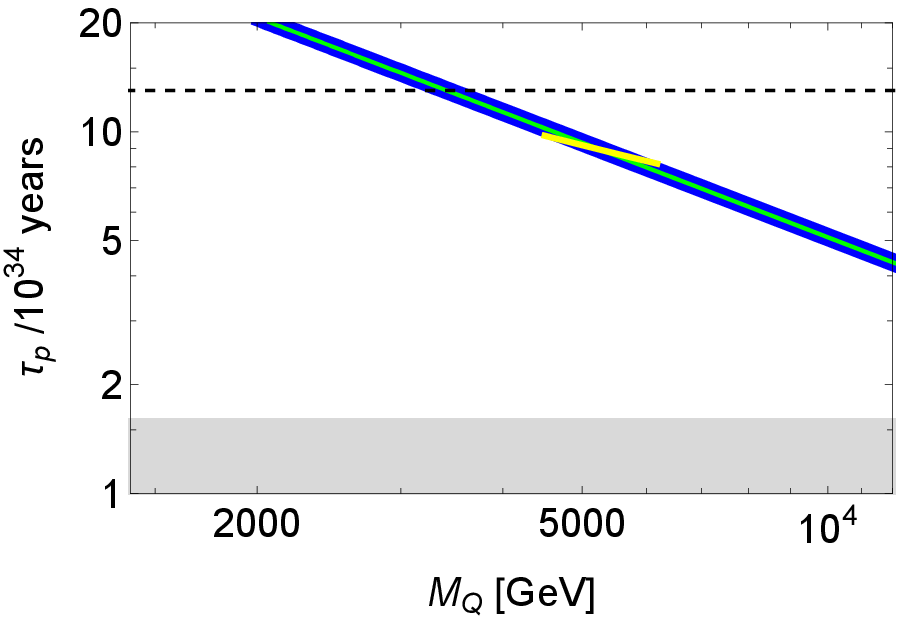}
\end{center}
\caption{
The blue (cyan) shaded region in the top left panel denote the range for the new quarks mass, $M_Q$, and the new lepton mass, $M_L$, to achieve the SM gauge couplings unification with an accuracy of $5\%$ ($1\%$) or less (see text for details). 
For reference, the yellow diagonal lines depict the contours for fixed $M_L = 4.0\times 10^{12}$ GeV for achieving the unification within $5\%$ accuracy.   For the mass values to the left of the solid black line, the SM Higgs potential is stabilized. 
For the new quarks and lepton masses identified in the top left panels, top right and bottom left panels, respectively, show the unification scale $M_{GUT}$ and the value of the unified coupling $\alpha_{GUT}$ as a function of $M_Q$. 
The bottom right panel shows the proton lifetime where the gray shaded region ($\tau_p \leq 1.6 \times 10^{35}$) in the bottom right panel denote the exclusion from Super-Kamiokande experiment.  The horizontal dashed line ( $\tau_p = 1.3 \times 10^{35}$) is the expected reach of the future Hyper-Kamiokande experiment. 
}
\label{fig:GCU2}
\end{figure}
In the top left panel of Fig.~\ref{fig:GCU2} the blue (cyan) shaded region denote the range for the new quarks mass, $M_Q$, and the new lepton mass, $M_L$, to achieve the SM gauge couplings unification with an accuracy of $5\%$ ($1\%$) or less. 
We define the accuracy as a percentage difference between the energy scales where the SM gauge couplings $\alpha_{1,2}$ and $\alpha_{2,3}$ are unified (see, for example, the running of couplings in Fig.~\ref{fig:GCU}). 
For reference, the yellow diagonal line depict the contours for fixed $M_L = 4.0\times 10^{12}$ GeV for achieving the unification with $5\%$ accuracy.   For the new fermion masses to the left of the solid black line, $\kappa _h(\mu) > 0$ for all values of $\mu$ and hence the SM Higgs potential is stabilized.

In the top right and bottom left panels of  Fig.~\ref{fig:GCU2} we show the unification scale $M_{GUT}$ and the value of the unified coupling $\alpha_{GUT}$ as a function of $M_Q$, respectively, for the new quarks and lepton masses identified in the top left panel of Fig.~\ref{fig:GCU2}. 
Here, $M_{GUT}$ is obtained by averaging the  energy scales satisfying $\alpha_1 =\alpha_2$ and $\alpha_2 =\alpha_3$. 
Requiring these two values to be within $5\%$ of each other, we find that the values of the SM gauge couplings at $M_{GUT}$ are well within a percent of each others value such that SM gauge couplings effectively unify at a single point at $M_{GUT}$.  
The bottom right panel shows the proton lifetime where the gray shaded region ($\tau_p \leq 1.6 \times 10^{35}$) denote the exclusion from Super-Kamiokande experiment. The horizontal dashed line ( $\tau_p = 1.3 \times 10^{35}$) is the expected reach of the future Hyper-Kamiokande experiment.

\section{Axion Dark Matter}
\label{sec:axionDM}
The relic abundance of axion DM is given by \cite{Kawasaki:2013ae}
\bea
\Omega_a h^2 &\simeq&  0.12 \; \left(\frac{\theta_a}{3.40 \times 10^{-3}}\right)^2 \left(\frac{f_a}{10^{16}\; {\rm GeV}}\right)^{1.19},  
\label{eq:DMab}
\eea
where $f_a=v_{PQ}/N_{DM}$ is the axion decay constant, $N_{DW}$ is the domain wall number and $\theta_a$ is the so-called misalignment angle. 
The observed DM relic abundance is $\Omega_a h^2 = 0.120 \pm 0.0012$ \cite{Aghanim:2018eyx}, and the axion decay constant is bounded from below by the measurement of the supernova SN 1987A pulse duration, $f_a  \gtrsim 4 \times 10^8$ GeV  \cite{Raffelt:2006cw}.

The axion/DM field fluctuation during inflation generates isocurvature density perturbations in the DM power spectrum, ${\cal P}_{\rm iso} =\left(\frac{H_{inf}}{\pi \theta_m  f_a }\right)^2 $, which is severely constrained by the Planck measurements \cite{Akrami:2018odb}
\bea
\beta_{\rm iso} \equiv \frac{{\cal P}_{\rm iso}(k_*)}{{\cal P}_{\rm iso}(k_*)+ {\cal P}_{\rm adi}(k_*)} < 0.038, 
\label{eq:isoPlanck}
\eea
where the adiabatic power spectrum ${\cal P}_{\rm adi} (k_*) \simeq 2.2 \times 10^{-9}$ with pivot scale $k_* = 0.05$ Mpc$^{-1}$ \cite{Aghanim:2018eyx}. 
We obtain  
\bea
\frac{H_{ inf}}{f_a} < 3.0 \times 10^{-5} \; \theta_a.   
\label{eq:Hiso1}
\eea 
In our model, $v_{PQ} = \sqrt{v_\Sigma^2 + v_S^2}$ and $N_{DW} = 3$ \cite{DiLuzio:2020wdo} such that $f_a = v_{PQ}/N_{DM} \simeq M_{GUT}$. 
From Eq.~(\ref{eq:DMab}), $\theta_a = 7.70\times 10^{-3}$ is fixed to reproduce the observed DM in the universe. 
Together with Eq.~(\ref{eq:DMab}), we obtain an upper bound $H_{inf} \lesssim 5.73 \times 10^8$ GeV. 
Therefore, the value of the Hubble parameter during inflation must be relatively low for the viability of the axion DM scenario that we have considered.  
For $f_a ={\cal O} (10^{16})$ GeV and higher, the the axion mass is ${\cal O} \lesssim (10^{-9})$ eV, which can be searched by the CASPEr experiment \cite{Budker:2013hfa}.

\section{Inflection-Point Inflation}
The inflaton potential that exhibits an approximate inflection-point around $\phi = M$ is given by  
\bea
V(\phi)\simeq V_0 +\ V_1 (\phi-M) +  \frac{V_2}{2} (\phi-M)^2 + \frac{V_3}{6} (\phi-M)^3, 
\label{eq:PExp}
\eea
   where $V_0 = V(M)$, $V_n \equiv  {\rm d}^{n}V/{\rm d} \phi^n |_{\phi =M}$, 
and $\phi = M$ is identified as the horizon exit scale corresponding to the pivot scale $k_* = 0.05$ Mpc$^{-1}$ used in Planck measurements \cite{Akrami:2018odb}.  
Requiring the inflationary predictions to be consistent with the Planck measurements \cite{Akrami:2018odb} of the curvature perturbation amplitude $\Delta_{\mathcal{R}}^2= 2.099 \times 10^{-9}$ and spectral index $n_s = 0.965$, 
$V_{1,2,3}$ can be expressed in terms of $V_0$, $M$ and the number of e-folds during the inflation $N$ as (see Ref.~\cite{Okada:2016ssd} for details)  
\bea
\frac{V_1}{M^3}&\simeq& 2.01 \times 10^3 \left(\frac{M}{M_P}\right)^3\left(\frac{V_0}{M^4}\right)^{3/2}, \nonumber \\
\frac{V_2}{M^2}&\simeq& -1.73 \times 10^{-2}  \left(\frac{M}{M_P} \right)^2 \left(\frac{V_0}{M^4}\right),  \nonumber \\
\frac{V_3}{M} &\simeq& 6.83 \times 10^{-7} \; \left( \frac{60}{N} \right)^2
   \left( \frac{M}{M_P} \right) \left( \frac{V_0}{M^4}   \right)^{1/2}. 
\label{eq:FEq-V1V2}
\eea 
For the remainder of this article, we set the e-folding number $N = 60$ to solve the horizon problem of big bang cosmology.

We identify $V(\phi)$ in Eq.~(\ref{eq:PExp}) with the RG improved $U(1)_X$ Higgs/inflaton potential 
\bea
V( \phi) =  \lambda_\phi (\phi) \left( \Phi^\dagger \Phi - \frac{v_X^2}{2}  \right)^2 \simeq \frac{1}{4} \lambda_\phi ( \phi)\; \phi^4,  
\label{eq:VEff1}
\eea
where $\lambda_\phi (\phi)$ is determined by solving the following RG equations:  
\bea
 \phi  \frac{d g}{d  \phi} &=& \frac{1}{16 \pi^2}  \left(\frac{264}{25}\right) g^3,         \nonumber\\
 \phi \frac{d Y_{i}}{d  \phi}   &=& \frac{1}{16 \pi^2}\left(Y_i^2+\frac{1}{2} \sum_{j=1}^3  Y_j^2-6 g^2 \right) Y_i,   
\nonumber\\
 \phi \frac{d \lambda_\phi}{d  \phi}  &=& \beta_{\lambda_\phi}.    
  \label{eq:RGEs}
\eea
Here, we have simplified the notation using $Y_i \equiv Y_M^i$ to denote the Majorana neutrino Yukawa couplings in Eq.~(\ref{eq:masses}), and the beta-function of ${\lambda_\phi}$ is given by
\bea
\beta_{\lambda_\phi} \!=\! \frac{1}{16 \pi^2}\! \left(\!20 \lambda_\phi^2  \!- 48\lambda_\phi  g^2\!+2 \lambda_\phi\sum_{i=1}^3 Y_i^2 
+96 g^4 - \sum_{i=1}^3 Y_i^4\!\right).
\label{eq:BGen}
\eea
Using the RG improved inflaton potential together with  the RG equation for $\lambda_\phi$, $V_{1,2,3}$ in Eq.~(\ref{eq:PExp}) may be expressed as 
\bea
\frac{V_1}{M^3}&=& \left.\frac{1}{4} (4 \lambda_\phi + \beta_{\lambda_\phi})\right|_{ \phi= M},\nonumber \\
\frac{V_2}{M^2}&=&  \left.\frac{1}{4} (12\lambda_\phi + 7\beta_{\lambda_\phi}+M \beta_{\lambda_\phi}^\prime)\right|_{ \phi= M}, \nonumber \\
\frac{V_3}{M}&=&  \left.\frac{1}{4} (24\lambda_\phi + 26\beta_{\lambda_\phi}+10M \beta_{\lambda_\phi}^\prime+M^2 \beta_{\lambda_\phi}^{\prime\prime})\right|_{ \phi= M}, 
\label{eq:ICons2}
\eea
where the prime denotes derivatives with respect to $\phi$.

Approximate inflection-point conditions at $M$, $V_1/M^3\simeq 0$ 
and 
$V_2/M^2\simeq 0$ 
yields 
$\beta_{\lambda_\phi} (M)\simeq -4\lambda_\phi(M)$ 
and 
$M\beta_{\lambda_\phi}^{\prime}(M) \simeq  16 \lambda_\phi (M)$. 
For concreteness, let us consider $Y_1 (M) < Y_{2,3}(M), g (M)$, a choice which will be justified shortly. 
For simplicity, we also set $Y_2(M) = Y_3(M)$. 
We later show that the inflection point conditions require $Y_{2,3}(M)$ and $g (M)$ to be of the same order and $\lambda_\phi (M) \propto g(M)^6$. 
Using this we can approximate $M^2 \beta_{\lambda_\phi}^{\prime\prime}(M) = - M \beta_{\lambda_\phi}^{\prime}(M)  +\phi \frac{{ d} }{{d}\phi} ( \phi \beta_{\lambda_\phi}^{\prime} ) |_{\phi = M} \simeq - M \beta_{\lambda_\phi}^{\prime}(M)$, 
where we have neglected $\phi \frac{{ d} }{{d}\phi} ( \phi \beta_{\lambda_\phi}^{\prime} ) |_{\phi = M}$ because it is a polynomial of degree $8$ in $g (M)$ and $Y_{2,3} (M)$, whereas $M \beta_{\lambda_\phi}^{\prime}(M)$ is a polynomial of degree $6$. 
This simplifies the last term in Eq.~(\ref{eq:ICons2}) to $V_3/M \simeq 16 \;\lambda_\phi(M)$. 
Together with the expression for $V_3/M$ in Eq.~(\ref{eq:FEq-V1V2}) and $V_0 \simeq (1/4) \lambda_\phi(M) M^4$, the quartic coupling is determined as  
\bea
\lambda_\phi(M)\simeq 4.56 \times 10^{-16} \left(\frac{M}{M_{P}}\right)^2. 
\label{eq:FEq} 
\eea
The Hubble parameter during inflation is given by  
\bea
H_{inf} &\simeq& \sqrt{\frac{V_0}{M_P^4}} \simeq 1.50\times 10^{10} \;{\rm GeV} \;\left(\frac{M}{M_P}\right)^2. 
\label{eq:FEqR} 
\eea
Substituting $H_{inf} \lesssim 5.73 \times 10^8$ GeV, the upper bound on Hubble parameter to solve the axion domain wall and isocurvature problems is expressed as 
\bea
\frac{M}{M_P} < 0.20 \left(\frac{f_a}{10^{16} \;{\rm GeV}}\right)^{0.135}.  
\label{eq:MBound}
\eea
For $H_{inf} \lesssim 10^9$ GeV, the inflationary prediction for the tensor-to-scalar ratio $r$ is tiny ($H_{inf} = 2.47 \times 10^{14} \; {\rm GeV} \sqrt{r}$).

To evaluate the masses of $Z^\prime$ gauge boson, Majorana neutrinos and inflaton, we now consider the low energy values of the relevant couplings.  
For concreteness, let us fix $Y_1 (M) = (\sqrt{2}/5)  g(M)$ by setting the mass ratio $m_{Z^\prime}/m_{N^1} = 10$ at $\phi = M$, and $Y_{2}(M) = Y_{3}(M)$. 
The inflection-point condition $\beta_{\lambda_\phi}(M) \simeq 0$ leads to 
\bea
Y_{2,3}(M)\simeq 2.63\;g(M). 
\label{eq:FEq3}
\eea
Evaluating the other inflection-point condition, $M\beta_{\lambda_\phi}^{\prime}(M) \simeq  16 \lambda_\phi (M)$, 
by using the RG equations in Eqs.~(\ref{eq:RGEs}) and~(\ref{eq:FEq3}), we obtain
\bea
\lambda_\phi(M) \simeq 2.56\times 10^{-3} \, g(M)^6. 
\label{eq:LandG}
\eea 
We note that the contributions of the new fermions to the beta-function of $g$ in Eq.~(\ref{eq:RGEs}) are key to obtaining $\lambda_\phi(M) > 0$ in Eq.~(\ref{eq:LandG}), which is essential for the stability\footnote{The authors in Ref.~\cite{Okada:2016ssd} have examined $U(1)_X$ Higgs Inflation without the vector-like fermion pairs and pointed out that $\lambda_\phi(M) < 0$  for $x_H = -4/5$.} of the $U(1)_X$ Higgs/inflaton potential. 
Equating the expressions for $\lambda_\phi(M)$ in Eqs.~(\ref{eq:RGEs}) and (\ref{eq:FEq}), we find
\bea
g(M)\simeq  7.50 \times 10^{-3} 
 \left(\frac{M}{M_{P}}\right)^{1/3}.
\label{eq:FEq2} 
\eea

\begin{figure}[t]
\begin{center}
\includegraphics[scale=0.68]{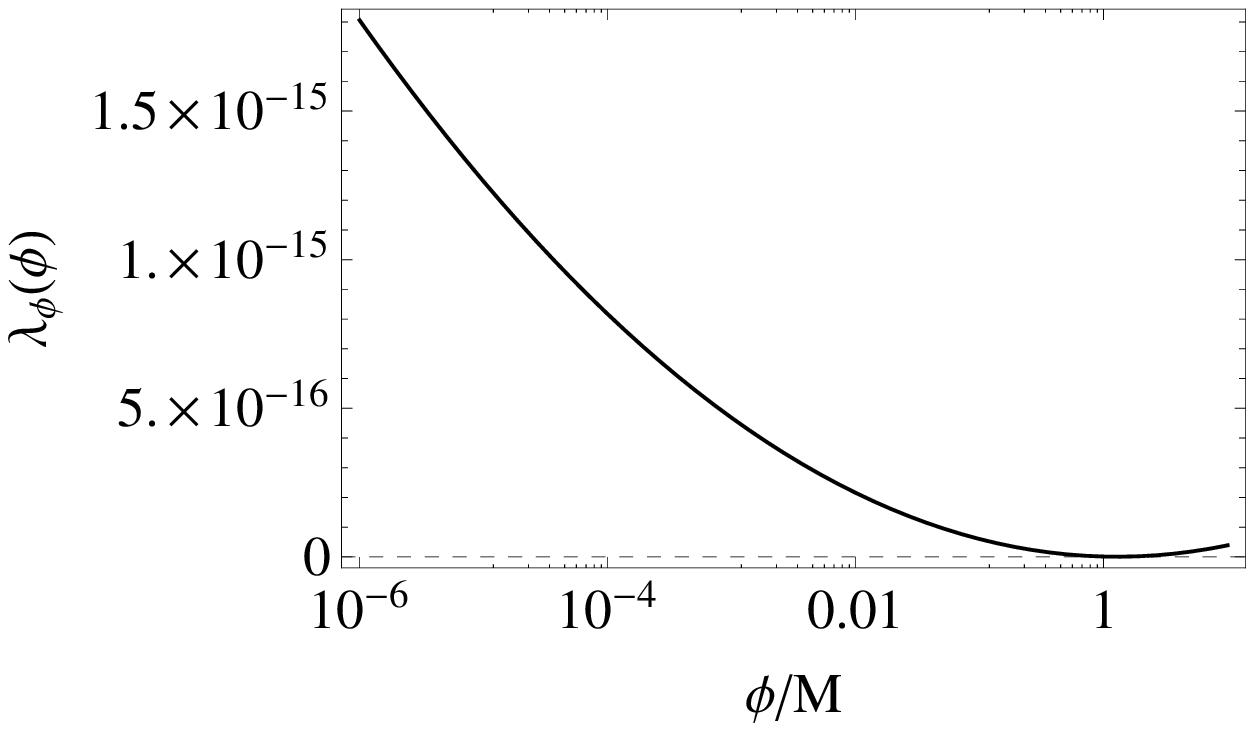}\;
\includegraphics[scale=0.64]{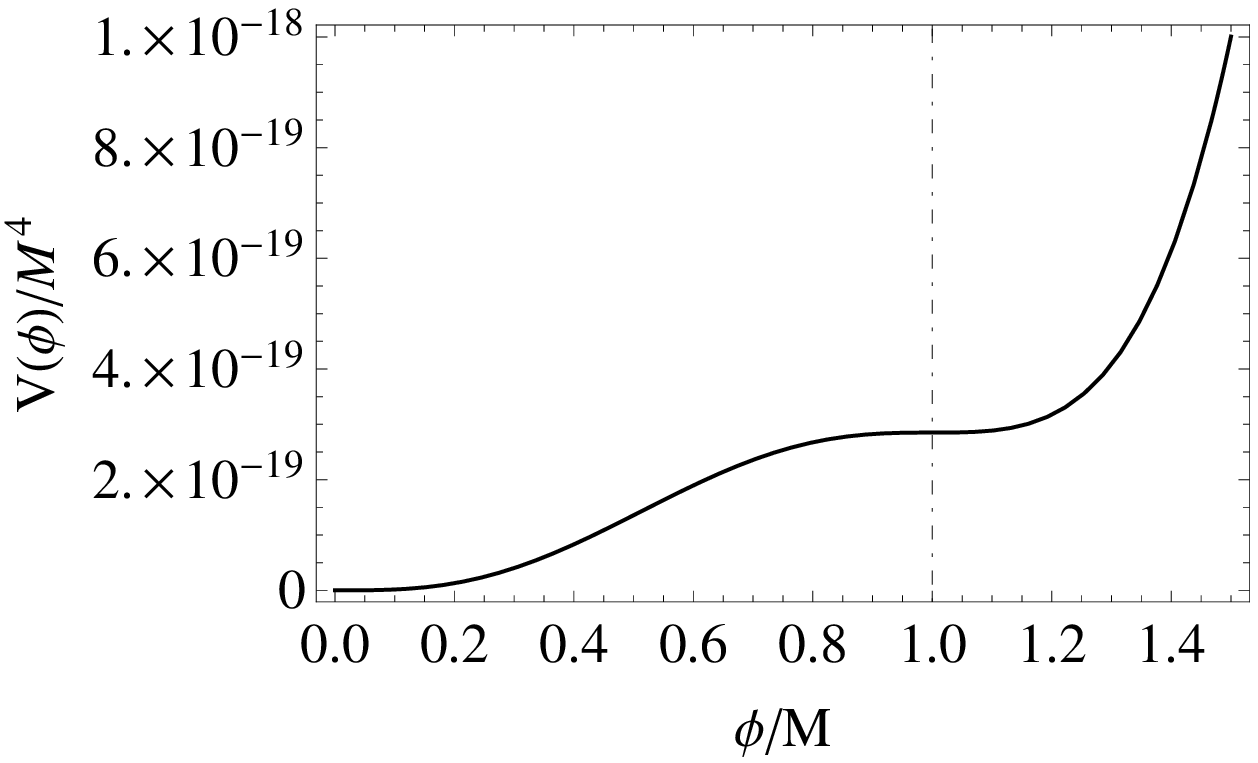}
\end{center}
\caption{
For fixed value of $M = 0.05 M_{P}$, the left panel shows the RG running of the inflaton quartic coupling $\lambda_\phi$ as a function of $ \phi/M$. 
The dashed horizontal line corresponds to $\lambda_\phi=0$. 
The right panel shows the RG improved effective inflaton potential with an approximate inflection-point at $ \phi \simeq M$ (vertical dashed-dotted line). 
}
\label{fig:IPI}
\end{figure}

Since the beta-function of the quartic coupling in Eq.~(\ref{eq:BGen}) is dominated by the gauge and Yukawa couplings, 
the RG equation for $\lambda_\phi$ can be solved analytically, and its value for $\phi \ll M$ can be estimated as \cite{Okada:2016ssd} 
\bea
\lambda_\phi( \phi) &=& \lambda_\phi(M) + 8 \lambda_\phi(M) \left(\ln \left[\frac{ \phi}{M}\right] \right)^2, 
\nonumber \\
&\simeq&  
3.81 \times 10^{-15}\left(\frac{M}{M_P}\right)^2\left(\ln \left[\frac{ \phi}{M}\right] \right)^2. 
\label{eq:Lmu} 
\eea
The masses of the inflaton, $Z^\prime$ boson and Majorana neutrinos in Eqs.~(\ref{eq:mass}) and~(\ref{eq:masses}), evaluated at $\phi =v_X$, are given by 
\bea
m_\phi &\simeq& 8.54 \times 10^{-8} v_X \left|{\rm ln}\left[\frac{v_X}{M}\right]\right|\left(\frac{M}{M_P}\right), 
\nonumber \\
m_{Z^\prime} &\simeq& 1.62 \times 10^{-2} 
 \left(\frac{M}{M_{P}}\right)^{\!\!1/3}, 
\nonumber \\
m_{N^{1}} &\simeq& \frac{m_{Z^\prime}}{10}, 
\nonumber \\
m_{N^{2,3}} &\simeq& 0.93 \; m_{Z^\prime}, 
\label{eq:ratiophiz} 
\eea 
where we have used $g(v_X) \simeq g(M) $ and $  Y_i (v_X) \simeq Y_i (M)$. 
Note that the new particle spectrum is determined by $v_X$ and $M$ in our model.

\section{Thermal Leptogenesis and Reheating}
\label{sec:Reheat}
To generate the observed baryon asymmetry we consider thermal leptogenesis \cite{Fukugita:1986hr}, which is the one of the simplest realization of the scenario in models with type-I seesaw mechanism. 
Since the Majorana neutrinos have non-degenerate masses, a successful thermal leptogenesis requires the lightest Majorana neutrino mass, $m_{N^1}>10^{9-10}$ GeV with reheat temperature $T_R > m_{N^1}$ \cite{Buchmuller:2002rq}. 
To prevent the $U(1)_X$ gauge interactions \cite{Iso:2010mv} and Yukawa interactions \cite{Dev:2017xry} from keeping the Majorana neutrinos in thermal equilibrium with the SM particles and suppressing the generation of lepton asymmetry, we require these processes to decouple before the temperature of the thermal plasma drops to $T \sim m_{N^1}$. 

For $m_{Z^\prime} > m_{N^{1}}$, the $Z^\prime$ mediated process, ${(N^c)^1} (N^c)^1 \to  Z^\prime \to \overline{f_{SM}} f_{SM} $, 
where $f_{SM}$ denote the SM fermions, is effectively a four-Fermi interaction. 
For $T > m_{N^1}$, the thermally-averaged cross section for this process is given by \cite{Okada:2016tci}
\bea
\langle \sigma v\rangle \simeq \frac{11}{1280\pi} \frac{T^2}{v_X^4}.  
\eea 
The annihilation/creation rate of $(N^c)^1$ in the thermal plasma $\Gamma (T) =  n_{eq}(T) \langle \sigma v\rangle$, where $n_{eq}(T) \simeq 2 T^3/\pi^2$  is the equilibrium number density. 
This process decouples at $T\sim m_{N^1}$ if $\left.\Gamma/H\right|_{T=m_{N^1}}< 1$, where $H(T) \simeq \pi T^2/M_P$ is the corresponding value of the Hubble parameter. It leads to a lower bound on $v_X$
\bea
v_{X} > 3.48 \times 10^{10} \;{\rm GeV} \left(\frac{m_{N^1}}{10^{9} \; {\rm GeV}}\right)^{3/4}. 
\label{eq:cond1}
\eea

The thermally averaged cross section for the process involving Yukawa interactions of  $N_R^{1}$, particularly, ${N_R}^{1} N_R^{1} \leftrightarrow  \phi \phi$,  with $m_{N^1}> m_\phi$ is approximated as \cite{Plumacher:1996kc}
\bea
   \langle \sigma v \rangle \simeq \frac{1}{4 \pi}   \frac{{m_{N^1}}^2}{v_{X}^4}. 
\label{eq:cs2}
\eea
Requiring $\Gamma/H < 1$ at $T=m_{N^1}$ to avoid the suppression of the generation of lepton asymmetry, we find 
\bea
v_{X} > 5.95 \times 10^{10} \;{\rm GeV} \left(\frac{m_{N^1}}{10^{9} \; {\rm GeV}}\right)^{3/4}, 
\label{eq:cond2}
\eea
which is slightly stronger than the lower bound obtained for the $Z^\prime$ mediated process in Eq.~(\ref{eq:cond1}).

Let us fix $M= 0.05 M_P$ to be our benchmark for consistency with the axion DM bound in Eq.~(\ref{eq:MBound}) with $F_a \simeq M_{GUT}$.
Together with $m_{N^1} = 10^{9} \; {\rm GeV}= m_{Z^\prime}/10$, we find  $v_X \simeq 1.70 \times 10^{12}$ GeV which is consistent with the above bound on $v_X$.
With these values, the mass of the remaining Majorana neutrinos $m_{N^{2,3}} \simeq 9.30 \times 10^9$ GeV, and the mass of the inflaton $m_\phi \simeq 8.10 \times 10^4$ GeV. 
As we have discussed earlier, for successful thermal leptogenesis, the reheat temperature ($T_R$) must satisfy $T_R>m_{N^1}$. 
Assuming an instantaneous decay of the inflaton field, 
the reheat temperature can be estimated as  
\bea
    T_R \simeq \left(\frac{90}{\pi^2 g_*}\right)^{1/4} \sqrt{\Gamma_\phi M_P},  
\label{eq:TR}
\eea
where $g_* \simeq 100$ and $\Gamma_\phi$ is the total decay width of the inflaton. 
To estimate $\Gamma_\phi$, we consider the following mixed quartic interaction between $\Phi$ and the SM doublet Higgs field $H$ in the scalar potential: 
\bea
V \supset  2 \lambda^\prime  \left(\Phi^\dagger \Phi\right) \left(H^\dagger H\right)
\supset 
 \lambda^\prime v_{X}  \phi H^\dagger H. 
\label{eq:InfPot}
\eea 
The decay width of $\phi$ is approximated as  
\bea
\Gamma_\phi \simeq  \frac{{\lambda^\prime}^2 v_X^2 }{8 \pi \, m_ \phi}, 
\label{eq:gamma2}
\eea
and the reheat temperature is given by 
\bea
T_R \simeq 10^{10} \;{\rm GeV} \left(\frac{\lambda^\prime}{9.86 \times 10^{-9}}\right).  
\label{Lambda}
\eea 
Hence, $T_R >m_{N^1}$ can be achieved with $\lambda^\prime \gtrsim 9.86\times10^{-9}$.

\section{Summary}
\label{sec:conc}
It is well-known that SM needs to be supplemented with new physics in order to address its inadequacies related to DM physics, neutrino masses and mixings, baryon asymmetry in the universe, cosmic inflation, and strong CP problem. We have proposed an extension of the SM which is based on $SU(5)$ grand unification that accounts for all of the above inadequacies.

Our model is based on $SU(5) \times U(1)_X \times U(1)_{PQ}$ symmetry, where the $U(1)_X$ gauge symmetry is the generalization of the $B-L$ symmetry, and $U(1)_{PQ}$ is the global Peccei-Quinn (PQ) symmetry.
It includes four fermion families in ${\bf {\overline 5}} + {\bf 10}$ representation of $SU(5)$, a mirror family in ${\bf { 5}} + {\bf {\overline 10}}$ representations, and three $SU(5)$ singlet three Majorana fermions. 
The $U(1)_X$ related anomalies cancel in the presence of the Majorana neutrinos.
The scalar sector includes four complex scalars, $\Sigma$, $S$, $H$ and $\Phi$.
The new fermions are essential for achieving a successful unification of the SM gauge couplings. 
We have shown that the SM gauge couplings unify at $M_{GUT} \simeq (6-9)\times 10^{15}$ GeV for a wide range of new fermion masses, and the proton lifetime $\tau_p$ is estimated to be well within the expected sensitivity of the future Hyper-Kamiokande experiment, $\tau_p \lesssim 1.3 \times 10^{35}$ years. 
The new fermions also stabilize the SM Higgs potential at high energies.
The $SU(5)$ adjoint scalar $\Sigma$ is also charged under the PQ symmetry, and hence the spontaneous breaking of the $SU(5)$ also triggers the breaking of the PQ symmetry, resulting in axion dark matter. 
The axion decay constant $f_a$ is of the same order as the $SU(5)$ symmetry breaking scale $M_{GUT}$ or somewhat greater.
For $f_a \sim 10^{16}$ GeV and higher, the mass of the axion DM mass is ${\cal O} (10^{-9})$ eV and smaller, which can be searched by the CASPEr experiment.
The value of the Hubble parameter during inflation must be low, $H_{inf} \lesssim 10^9 $ GeV,  in order to successfully resolve the axion domain wall, axion DM isocurvature, and $SU(5)$ monopole problems. 
With the identification of the $U(1)_X$ Higgs field with the inflaton field, we have implemented the low-scale inflection-point inflation which is capable of realizing the desired value for $H_{inf}$.
The new fermions are also essential for the phenomenological viability of both the axion DM and inflation scenarios. 
We have also shown that the inflaton decay after the end of inflation can reheat the universe to a sufficiently high temperature such that the Majorana fermions generate the observed baryon asymmetry in the universe via  leptogenesis.

\section{Acknowledgements}
This work is supported in part by the United States Department of Energy grant DE-SC0012447 (N.~Okada) 
and DE-SC0013880 (D.~Raut and Q.~Shafi).

\section*{Appendix 1}
\label{app1}
Expanding the Higgs potential in Eq.~(\ref{eq:SSPot}) around the potential minimum along with the stationary conditions in Eq.~(\ref{eq:mincond}), we obtain the following non-zero mass eigenvalues after the diagonalization of the scalar mass matrix:
\bea
m_1^{(1)} &=& \sqrt{\frac{\left(v_S^2+v_{\Sigma }^2\right) \left(120 \kappa_{3} v_S+\sqrt{30} \kappa_{1} v_{\Sigma }\right)}{120 v_S}}, 
\label{eq:m1}
\\
m_2^{(3)} &=&\sqrt{\frac{1}{120} \left(-3 \sqrt{30} \kappa_1 v_S v_{\Sigma }+120 \kappa_3 v_S^2+4 \left(-30 \kappa _2+\kappa _3+\kappa _4\right) v_{\Sigma }^2\right)}, 
\label{eq:m2}
\\
m_3^{(8)} &=& \sqrt{\frac{1}{120} \left(7 \sqrt{30} \kappa_1 v_S v_{\Sigma }+120 \kappa_3 v_S^2-6 \left(20 \kappa _2+\kappa _3+\kappa _4\right) v_{\Sigma }^2\right)}, 
\label{eq:m3}
\\
m_4^{(8)} &=&\sqrt{\frac{1}{24} v_{\Sigma } \left(2 \left(\kappa _3+\kappa _4\right) v_{\Sigma }+3 \sqrt{30} \kappa_1 v_S\right)}, 
\label{eq:m4}
\\
m_5^{(3)} &=&\sqrt{\frac{1}{24} v_{\Sigma } \left(8 \left(\kappa _3+\kappa _4\right) v_{\Sigma }-3 \sqrt{30} \kappa_1 v_S\right)}, 
\label{eq:m5}
\\
m_6^{(12)} &=&\sqrt{\frac{1}{60} \left(\sqrt{30} \kappa_1 v_S v_{\Sigma }+60 \kappa_3 v_S^2+\left(-60 \kappa _2+12 \kappa _3-13 \kappa _4\right) v_{\Sigma }^2\right)}, 
\label{eq:m6}
\\
m_7^{(1)} &=& \sqrt{\frac{1}{240 v_S}\left(f_1 + \sqrt{f_1^2 - 16 v_S v_\Sigma \times f_2}\right)},
\label{eq:m7}
\\
m_8^{(1)} &=& \sqrt{\frac{1}{240 v_S}\left(f_1 - \sqrt{f_1^2- 16 v_S v_\Sigma \times f_2}\right)}, 
\label{eq:m8}
\eea
where the terms in exponents for each $m_i$ mass eigenvalues indicate the degeneracy of the masses and 
\bea
f_1 &=& v_{\Sigma } \left(4 \left(30 \kappa _1+30 \kappa _2+7 \left(\kappa _3+\kappa _4\right)\right) v_S v_{\Sigma }-\sqrt{30} \kappa_1 \left(3 v_S^2-v_{\Sigma }^2\right)\right)+120 \kappa _S v_S^3, 
\nonumber \\
f_2 &=& -\sqrt{30} \kappa_1 \left(-90 \left(\kappa_2-2 \kappa_3\right) v_S^2 v_{\Sigma }^2+90 \kappa _S v_S^4-\left(30 \kappa _1+30 \kappa _2+7 \left(\kappa _3+\kappa _4\right)\right) v_{\Sigma }^4\right)
-90 \kappa_1^2 v_S v_{\Sigma }^3
 \nonumber \\ && 
+60 v_S^3 v_{\Sigma } \left(2 \left(30 \kappa _1+30 \kappa _2+7 \left(\kappa _3+\kappa _4\right)\right) \kappa _S-15 \left(\kappa_2-2 \kappa_3\right){}^2\right). 
\nonumber
\eea

For simplicity, we fix $v_{\Sigma }=  \sqrt{30} \kappa_{1} v_S$ and $\lambda_3 = \lambda_4 $ in the following analysis.  
Requiring positive mass eigenvalues for our benchmark parameters, 
Eq.~(\ref{eq:m1}) leads to $\kappa_3 > -\kappa_{1}^2/4$ while Eqs.~(\ref{eq:m4}) and (\ref{eq:m5}) leads to $\lambda_4 > 3/16$. 
Defining $\kappa_3 = (-1+r) \kappa_{1}^2/4$ with $r>0$ and requiring $m_i^2 (i = 1, 2, .., 6) >0$ leads to constraint on other coupling parameters. 
For $\frac{3}{16} < \lambda_4 \leq \frac{1}{2} $
\bea
\lambda_2 < \frac{1}{120} \left(-4 + r + 8 \lambda_4\right), 
\eea
and for $\lambda_4 > \frac{1}{2} $
\bea
\lambda_2 < \frac{1}{120} \left(-6 + r + 12 \lambda_4\right). 
\eea
One needs to explicitly evaluate masses to ensure that $m_{7,8}^2 >0$. 
For example, with $r = 0.5$ and $\kappa_{1,2} = \lambda_S=  \lambda_1 = 0.3$, $\kappa_{3}  = -0.011$, $\lambda_2 = -0.049$, $\lambda_3 = \lambda_4  = 0.375$, the scalar mass spectrum is given by
\bea
m_1^{(1)} &=& 0.20 \;v_S, 
\qquad 
m_2^{(3)} = 0.35 \;v_S
\qquad
m_3^{(8)} =  0.42\;v_S
\qquad
m_4^{(8)} = 0.71\;v_S
\nonumber \\
m_5^{(3)} &=& 0.58 \;v_S
\qquad
m_6^{(12)} = 0.39 \;v_S
\qquad
m_7^{(1)} = 0.57 \;v_S
\qquad
m_8^{(1)} = 1.1 \;v_S. 
\eea

\section*{Appendix 2}
\label{app2}
The RG equations for the SM gauge couplings ($g_{1,2,3}$), Yukawa coupling ($y_t$), and the Higgs coupling ($\lambda_h$) in the presence of the new quarks ($Q$ and $U^c$) and the new lepton ($L$) and their vector-like partners are given below. 
In the following we consider the contribution of the SM particles and the new fermions to the RG equations at 1-loop order in perturbation while at the 2-loop order we only consider contributions of the SM particles.  
\bea
 \mu  \frac{{\rm d} g_1}{{\rm d}\mu}  = g_1^3 \left( \beta_{g_1}^{\rm 1-loop} +\beta_{g_1}^{\rm 2-loop} ({\rm SM})\right)
\eea
where 
\bea
\beta_{g_1}^{\rm 1-loop}  &=& \frac{1}{16\pi^2} \left(\frac{41}{10} + \frac{2}{5}\times \rm{\theta(\mu -M_Q)}+  \frac{2}{5} \times \rm{\theta(\mu -M_L)}\right),
\nonumber \\
\beta_{g_1}^{\rm 2-loop}  ({\rm SM}) &=& \left( \frac{1}{16\pi^2}\right)^2 
\left(
\frac{199}{50} g_1^2+\frac{27}{10} g_2^2+\frac{44}{5} g_3^2 
-\frac{17}{10} y_t^2 
\right),
\nonumber 
\eea
and $M_Q$ is the common mass (for simplicity) of the new vector-like quarks and $M_L$ is the mass of the new vector-like lepton doublet. 
\bea
 \mu  \frac{{\rm d} g_2}{{\rm d}\mu}  = g_2^3 \left( \beta_{g_2}^{\rm 1-loop} +\beta_{g_2}^{\rm 2-loop}  ({\rm SM})\right),  
\eea
where 
\bea
\beta_{g_2}^{\rm 1-loop}  &=& \frac{1}{16\pi^2} \left(-\frac{19}{6} + 2\times \rm{\theta(\mu -M_Q)}+  \frac{2}{3} \times \rm{\theta(\mu -M_L)}\right),
\nonumber \\
\beta_{g_2}^{\rm 2-loop}  ({\rm SM}) & =& \left( \frac{1}{16\pi^2}\right)^2 
\left(
\frac{9}{10} g_1^2+\frac{35}{6} g_2^2+ 12 g_3^2 
-\frac{3}{2} y_t^2 
\right). 
\nonumber 
\eea

\bea
 \mu  \frac{{\rm d} g_3}{{\rm d}\mu}  = g_2^3 \left( \beta_{g_3}^{\rm 1-loop} +\beta_{g_3}^{\rm 2-loop} ({\rm SM})\right), 
\eea
where 
\bea
\beta_{g_3}^{\rm 1-loop}  &=& \frac{1}{16\pi^2} 
\left(-7 + 2\times \rm{\theta(\mu -M_Q)} \right), 
\nonumber \\
\beta_{g_3}^{\rm 2-loop} ({\rm SM})  &=& \left( \frac{1}{16\pi^2}\right)^2 
\left(
\frac{11}{10} g_1^2+\frac{9}{2} g_2^2 - 26 g_3^2  - 2 y_t^2 
\right). 
\nonumber 
\eea
\bea
 \mu  \frac{{\rm d} y_t}{{\rm d}\mu}  = y_t \left( \beta_{y_t}^{\rm 1-loop} +\beta_{y_t}^{\rm 2-loop}  ({\rm SM})\right), 
\eea
where 
\bea
\beta_{y_t}^{\rm 1-loop}  &=& \frac{1}{16\pi^2} \left(-\frac{17}{20} g_1^2-\frac{9}{4} g_2^2 - 8 g_3^2 +\frac{9}{2} y_t^2\right),
\nonumber \\
\beta_{y_t}^{\rm 2-loop}  ({\rm SM}) &=& \left( \frac{1}{16\pi^2}\right)^2 
\left(
\frac{11}{10} g_1^2+\frac{9}{2} g_2^2 - 26 g_3^2 + \left(\frac{1}{5} g_1^2 + 3 g_2^3 + 38 g_3^2\right) - 2 y_t^2 
\right). 
\nonumber 
\eea

\bea
 \mu  \frac{{\rm d} \kappa}{{\rm d}\mu}  = y_t \left( \beta_{\kappa}^{\rm 1-loop} +\beta_{\kappa}^{\rm 2-loop} ({\rm SM}) \right), 
\eea
where 
\bea
\beta_{\kappa}^{\rm 1-loop}  &=& \frac{1}{16\pi^2} 
\left(
12 \kappa^2 -\left(\frac{9}{5} g_1^2 +9g_2^2\right)\kappa
+\frac{9}{4}\left(\frac{3}{25}g_1^4 + \frac{2}{5} g_1^2 g_2^2 +g_2^4\right)
+12 y_t^2 \kappa -12 y_t^4 
\right),
\nonumber \\
\beta_{\kappa}^{\rm 2-loop}  ({\rm SM})  &=& 
\left(\frac{1}{16\pi^2}\right)^2
\left[
-78\kappa^3
+18\left( \frac{3}{5} g_1^2+3g_2^2\right)\kappa^2
- \left(\frac{73}{8} g_2^4 -\frac{117}{20} g_1^2g_2^2 -\frac{1887}{200} g_1^4\right) \kappa 
\right.
\nonumber \\
 &&\left.3\kappa y_t^4 +\frac{305}{8} g_2^6
 -\frac{289}{40} g_1^2g_2^4  
 -\frac{1677} {200} g_1^4 g_2^2
 -\frac{3411}{1000} g_1^6
 -64 g_3^2 y_t^4
 -\frac{16}{5} g_1^2y_t^4
 -\frac{9}{2} g_2^4y_t^2 \right.
\nonumber \\
&& \left.+10\kappa \left(\frac{17}{20} g_1^2 +\frac{9}{4} g_2^2 +8g_3^2\right)y_t^2
 -\frac{3}{5}g_1^2 \left(\frac{57}{10}g_1^2 -21g_2^2\right)y_t^2
 -72\kappa^2 y_t^2
 +60y_t^6
 \right]. 
\nonumber 
\eea


\end{document}